\documentclass[10pt,twocolumn]{article}
\usepackage[a4paper,margin=1in]{geometry}
\usepackage{amsmath}
\usepackage{hyperref}
\usepackage{svg}
\usepackage{graphicx}
\usepackage{tabularx}
\usepackage{microtype}

\hypersetup{
  colorlinks=false,
  pdfborder={0 0 0}
}

\title{Towards Performance-Aware Allocation for \\ Accelerated Machine Learning on GPU-SSD Systems}
\author{
  Ayush Gundawar \\
  \small Georgia Institute of Technology \\
  \small gundawar@gatech.edu
  \and
  Euijun Chung \\
  \small Georgia Institute of Technology \\
  \small euijun@gatech.edu
  \and
  Hyesoon Kim \\
  \small Georgia Institute of Technology \\
  \small hyesoon@cc.gatech.edu
}
\date{}

\begin{document}
\maketitle

\begin{abstract}
    \noindent
    The exponential growth of data-intensive machine learning workloads has exposed significant limitations in conventional GPU-accelerated systems, especially when processing datasets exceeding GPU DRAM capacity. We propose MQMS, an augmented in-storage GPU architecture and simulator that is aware of internal SSD states and operations, enabling intelligent scheduling and address allocation to overcome performance bottlenecks caused by CPU-mediated data access patterns. MQMS introduces dynamic address allocation to maximize internal parallelism and fine-grained address mapping to efficiently handle small I/O requests without incurring read-modify-write overheads. Through extensive evaluations on workloads ranging from large language model inference to classical machine learning algorithms, MQMS demonstrates orders-of-magnitude improvements in I/O request throughput, device response time, and simulation end time compared to existing simulators.
\end{abstract}

\section{Introduction}
The exponential growth in machine learning workload complexity and dataset sizes has exposed critical limitations in conventional GPU-accelerated systems, particularly for data-intensive applications such as graph neural networks (GNNs) and large-scale recommender systems. Current GPU-accelerated systems rely heavily on CPU-mediated data access patterns, creating a significant performance bottleneck due to the I/O and synchronization overhead stemming from data propagation over PCI-e links between components in the interaction model \cite{Liu_2023}. This architectural limitation becomes particularly pronounced when processing datasets exceeding GPU DRAM capacity, with data propagation overhead accounting for more than 80\% of total processing latency in GNN applications \cite{Liu_2023}. Current storage interfaces struggle to accommodate these unpredictable access patterns and dense I/O request bursts efficiently.

Direct GPU-SSD systems have emerged as a promising architectural solution, offering the potential to bypass CPU mediation in storage operations and leverage the full parallel processing capabilities of modern SSDs. However, current implementations face several critical challenges that prevent optimal performance. Enterprise SSDs employ sophisticated internal architectures with multiple channels, complex garbage collection mechanisms, and dynamic wear-leveling algorithms \cite{enterprise_design}. Yet, these components are typically treated as a black box by GPU interfaces in GPU-SSD systems. This abstraction leads to missed optimization opportunities, manifesting in reduced bandwidth utilization, poor data locality, and suboptimal channel traffic distribution \cite{Hong_Lee_Kim_2022}. Additionally, the smaller access granularity required by many machine learning workloads often conflicts with the underlying SSD architecture's optimal access patterns \cite{Liu_2023}.

To address these fundamental limitations, we propose a novel in-storage GPU architecture that maintains awareness of internal SSD states and operations. Our approach introduces mechanisms for intelligent scheduling and address allocation while maintaining full compatibility with host CPU I/O operations. We evaluate our proposed architecture through experimentation using an augmented version of the MQSim-MacSim simulator, called MQMS, which combines SSD emulation capabilities with cycle-accurate GPU simulation. Through MQMS, we assess system performance across multiple metrics---including I/O per second (IOPS) and device response time---on various workloads ranging from LLMs to classical learning algorithms.

\section{Enterprise-Grade SSDs}
Enterprise SSDs must satisfy stringent I/O requirements for performance-driven data-intensive applications, such as big data analytics services. These workloads demand high IOPS for parallel request processing and sustainable QoS to maintain consistent application performance. Meeting these requirements necessitates additional hardware resources and advanced internal functionalities compared to client SSDs \cite{Hong_Lee_Kim_2022}.

Performance analysis of current SSD simulator submodules reveals significant limitations in modeling enterprise-grade performance characteristics. Testing with 4KB random request workloads demonstrates that client SSD simulators, even when configured with enterprise-class parameters such as channel counts, die size, and page size, achieve only a fraction of real enterprise SSD performance \cite{Hong_Lee_Kim_2022}. While the Samsung PM9A3 enterprise SSD datasheet shows near-linear IOPS scaling with I/O queue size until saturation, simulators like SimpleSSD and MQSim scale at an asymptotic, non-linear rate, performing an order of magnitude worse \cite{pm9a3}.

While current simulators allow configuring physical specifications like die/block/page counts and read/write latencies, their IOPS is limited by the lack of resource management used in enterprise storage systems \cite{enterprise_design}. To address this, we implement two key optimizations inspired by enterprise SSD design in MQMS, leveraging core primitives like NVMe interface support inherited from the SSD simulator submodule, MQSim \cite{Tavakkol_Gomez-Luna_Sadrosadati_Ghose_Mutlu_2022}.

\subsection{Dynamic Address Allocation}
Enterprise SSDs maximize internal parallelism through numerous units such as channels, dies, and planes \cite{Tavakkol_Arjomand_Sarbazi-Azad_2014}. However, existing SSD simulators fail to fully exploit these units, particularly plane-level parallelism, resulting in reduced I/O request throughput potential \cite{Tavakkol_Mehrvarzy_Arjomand_Sarbazi-Azad_2016}.

The primary cause of this inefficiency is the \textit{static behavior} of the address allocation strategy employed by these simulators. Physical addresses for write I/O data are determined based on logical addresses using fixed rules. This approach limits opportunities to utilize underlying planes in parallel due to constraints associated with plane-level parallelism \cite{Hong_Lee_Kim_2022}. For instance, when physical addresses are statically assigned, multiple write I/O requests that could otherwise be processed in parallel may be forced to wait because they are mapped to the same plane \cite{Tavakkol_Mehrvarzy_Arjomand_Sarbazi-Azad_2016}. This results in I/O requests stalling while available planes remain idle, leading to underutilization of the SSD's potential parallelism.

\begin{figure}
    \centering
    \includesvg[width=\linewidth]{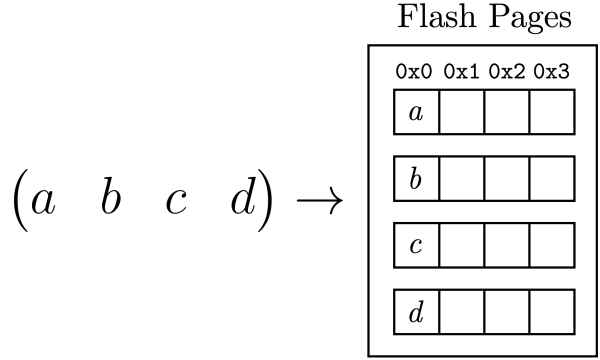}
    \caption{Dynamic Address Allocation}
    \label{fig:daa}
\end{figure}

To address this issue, modern enterprise SSDs employ \textit{dynamic} address allocation, allowing write I/O data to be stored anywhere by dynamically assigning physical addresses. This strategy enables the allocation of the same physical address—across different planes—to multiple arbitrary write I/O requests, thereby creating opportunities to exploit plane-level parallelism \cite{Tavakkol_Arjomand_Sarbazi-Azad_2014}. Figure~\ref{fig:daa} depicts an example where a vector in memory has its four contiguous elements dynamically allocated to the same address across four flash pages in parallel. By dynamically distributing write requests across all available planes, the SSD controller scales I/O request throughput as $\mathcal{O}(\min(n, p))$, where $n$ is the write request count and $p$ is the plane count.

It is important to consider that dynamic address allocation trades off plane-level data locality for higher request throughput. By allocating data across multiple planes, workloads that rely on plane-level locality may experience performance penalties. This trade-off is negligible for workloads with high data isolation. In the context of in-storage GPU systems, where workloads tend to be highly concurrent, maximizing SSD request throughput is generally a larger factor in minimizing workload latency \cite{Tavakkol_Mehrvarzy_Arjomand_Sarbazi-Azad_2016}.

We have implemented dynamic address allocation in MQMS, aiming to maximize internal parallelism for data-intensive workloads. Our experiments indicate that this approach outperforms restricted dynamic allocation methods in terms of device IOPS and response times for tested enterprise workloads.

\subsection{Fine-Grained Address Mapping}

Modern enterprise SSDs increasingly handle large numbers of small I/O requests, a trend exacerbated by the growth in flash page sizes (up to 16 KB) aimed at improving disk capacity and bandwidth \cite{enterprise_design}. However, current SSD simulators inefficiently process small I/O requests, generating redundant read and write operations that result in significantly poorer performance metrics relative to real devices.

\begin{figure}
    \centering
    \includesvg[width=0.9\linewidth]{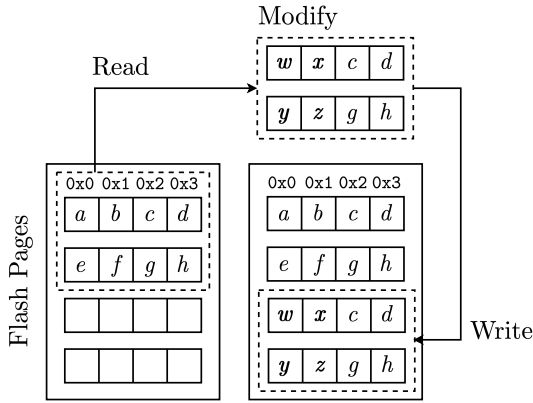}
    \caption{Coarse-Grained Mapping}
    \label{fig:cgm}
\end{figure}

This inefficiency stems from the coarse-grained address mapping schemes typically used in these simulators, where logical and physical addresses are mapped at the flash page level \cite{Tavakkol_Gomez-Luna_Sadrosadati_Ghose_Mutlu_2022}. Under page-level mapping, servicing a small write request involves three steps: reading the entire page containing the data to be updated, modifying the specific portion, and then writing back the entire page---a process of transactions known as read-modify-write (RMW) operations. Figure~\ref{fig:cgm} illustrates a scenario involving coarse-grained address mapping, where servicing four small write requests ($w$, $x$, $y$, and $z$) requires reading two flash pages, modifying their contents, writing the modified pages back, and marking the original pages as invalid.

\begin{figure}
    \centering
    \includesvg[width=0.9\linewidth]{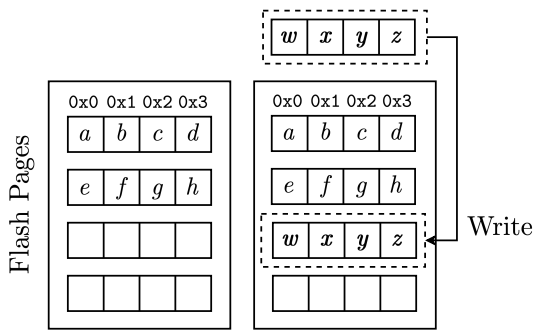}
    \caption{Fine-Grained Mapping}
    \label{fig:fgm}
\end{figure}

To address this issue, modern enterprise SSDs implement fine-grained address mapping, where logical and physical addresses are mapped at sub-page sizes, such as sectors or sub-sectors \cite{enterprise_design}. This method allows small write requests to be serviced immediately by writing only the new data and invalidating the corresponding old data, without reading or writing the unaffected portions of the page. Figure~\ref{fig:fgm} illustrates an example of fine-grained address mapping, where servicing four small write requests ($w$, $x$, $y$, and $z$) involves writing only a single flash page and invalidating the corresponding original data.

While fine-grained mapping increases the size of the mapping table and the frequency of table accesses, especially for large I/O requests, these overheads are mitigated in modern enterprise SSDs equipped with large internal DRAM caches that store the entire mapping table. The additional memory and computational overhead are generally negligible compared to the performance benefits gained from reduced RMW operations for small I/O requests \cite{Hong_Lee_Kim_2022}.

\section{Workload Evaluation}
\subsection{Kernel Sampling}
To evaluate MQMS against the baseline MQSim-MacSim simulator, we require representative traces from large-scale machine learning workloads. Modern machine learning models exhibit repeated kernel patterns derived from their block structure. For instance, ResNet-50 contains 48 identical convolutional layers, while transformer models iterate through self-attention and fully connected layer blocks \cite{he_2015}. Analysis reveals that these kernels exhibit independently and identically distributed execution times with negligible inter-kernel cache dependency, as evidenced by consistent L1 and L2 hit rates under cache flush conditions \cite{Chung_Na_Kim_2024}. This characteristic enables statistical kernel sampling for trace generation.

We leverage Allegro to generate workload traces for LLM inference, clustering GPU kernels by name, grid size, and block size before applying 1-D k-means clustering ($k=2$) recursively \cite{Chung_Na_Kim_2024}. The clustering process uses the central limit theorem to determine optimal group sizes, splitting groups until homogeneous execution time distributions within each cluster are achieved. For any kernel group $K$ with $N$ kernels, mean $\mu$ and variance $\sigma^2$, the sampled execution times $X_1...X_m$ converge to a normal distribution with mean $\mu$ and variance $\sigma^2/m$ as $m$ approaches infinity.

Per-group sampling uses $m_{min}$ kernels, where $m_{min}$ is derived from execution time variance relative to the mean to maintain error bounds. With kernel groups $K_i$ having $N_i$ kernels and sampled mean execution times $\bar{X_i}$ following normal distributions $N(\mu_i,\sigma_i^2)$, the total predicted execution time $Y = \sum N_i\bar{X_i}$ maintains specified error bounds $\epsilon$ with 95\% confidence \cite{Chung_Na_Kim_2024}.

We have implemented this sampling algorithm into MQMS's trace generation tool for GPU simulation, which creates SASS-Assembly traces for simulating NVIDIA GPU workloads. The algorithm dramatically reduces trace sizes while maintaining the essential workload characteristics needed for accurate comparative analysis. Implementation-wise, Allegro requires just a single upfront cost---while the initial hardware profiling and sampling do add some processing time, these steps occur only once during trace generation. After this preprocessing phase, the actual simulation runs with minimal additional performance overhead \cite{Chung_Na_Kim_2024}.

\subsection{Evaluation Results}
To evaluate the performance characteristics of MQMS against the baseline MQSim-MacSim simulator, we conducted experiments using representative workload traces generated from three data-intensive LLM inference workloads. LLM inference presents an ideal test case for storage system evaluation due to its data-intensive nature and the large-scale memory access patterns required to load model weights and process inputs. Table \ref{table:llm_workloads} provides descriptions of these sampled workloads.

\begin{table}
    \centering
    \begin{tabularx}{\columnwidth}{ll>{\raggedright\arraybackslash}X}
        \hline
        \textbf{Name}                                            & \textbf{Kernels} & \textbf{Description}                                         \\
        \hline
        BERT \cite{Bhargava_Drozd_Rogers_2020}                   & 1,858,800        & Classification of 10K premise \& hypothesis pairs            \\
        GPT-2 \cite{Radford_Wu_Child_Luan_Amodei_Sutskever_2019} & 34,981,000       & Generation of 1K sentences, each with a length of 100 tokens \\
        ResNet-50 \cite{He_Zhang_Ren_Sun_2015}                   & 2,812,741        & Classification of 13.4K ImageNet samples                     \\
        \hline
    \end{tabularx}
    \caption{Large-Scale Workloads}
    \label{table:llm_workloads}
\end{table}

Both simulators were evaluated under identical configurations tailored to emulate enterprise SSDs \cite{Hong_Lee_Kim_2022}. Key parameters, such as channel count, chips per channel, planes per die, and page size, were set to reflect enterprise SSD specifications. Timing configurations, such as read and write latencies, were adjusted to match specification performance as well. Both simulators were set to use the round-robin scheduling policy for GPU kernel execution and the CWDP (Channel-Way-Die-Plane) scheme for SSD page allocation \cite{Kim_2023}.

\begin{figure}
    \centering
    \includegraphics[width=\linewidth]{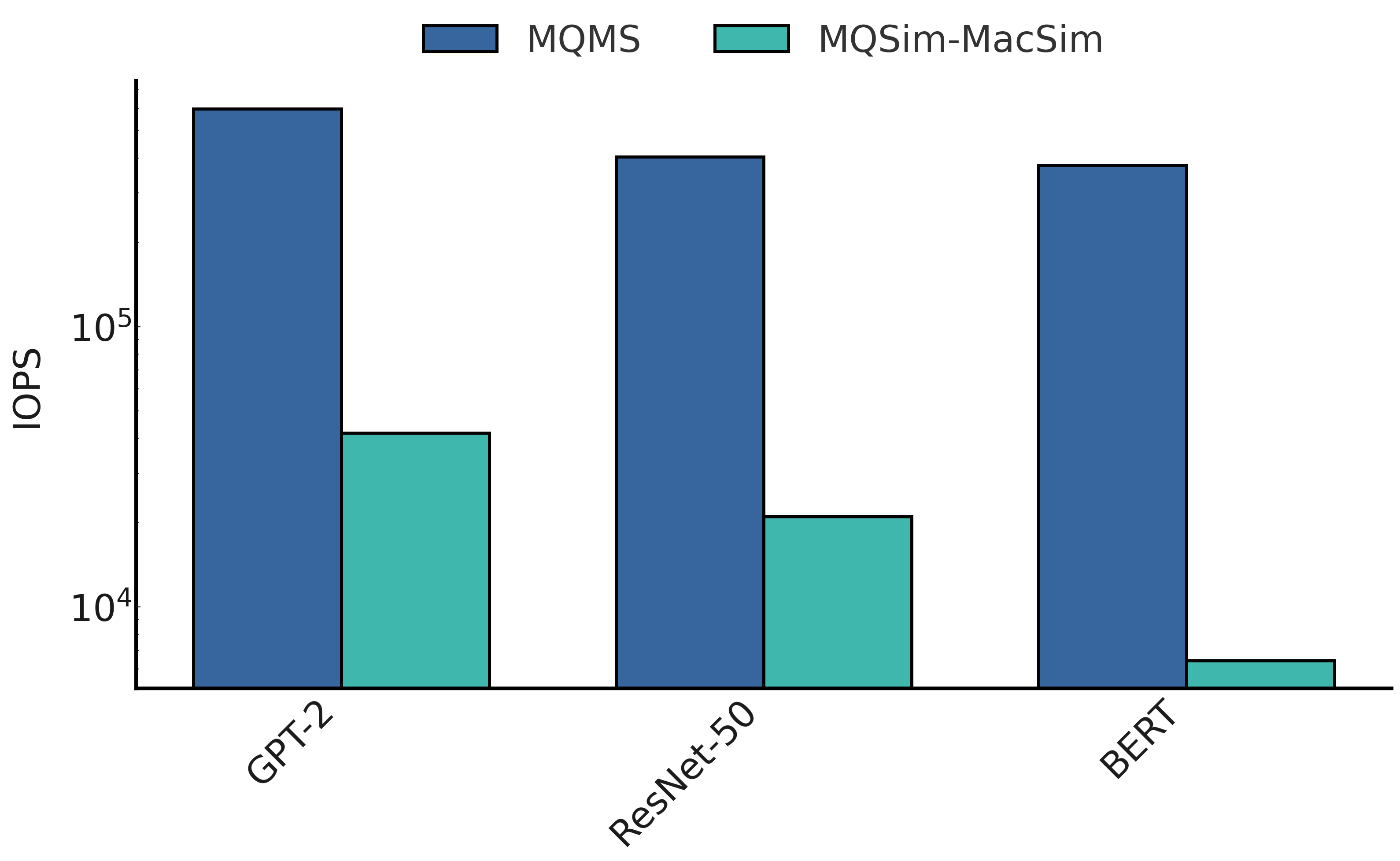}
    \caption{IOPS by Workload}
    \label{fig:llm_iops}
\end{figure}

\begin{figure}
    \centering
    \includegraphics[width=\linewidth]{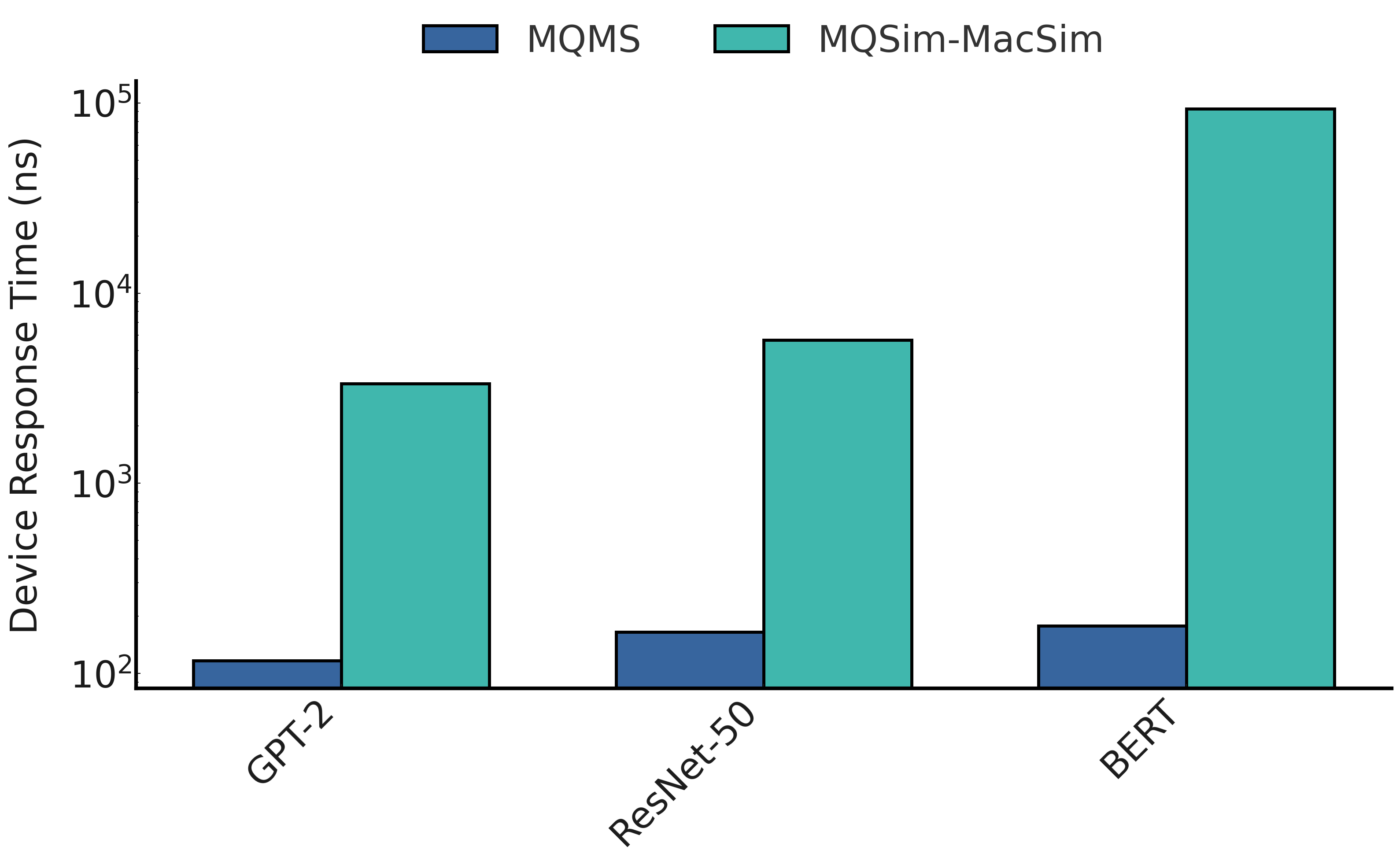}
    \caption{Device Response Time by Workload}
    \label{fig:llm_resp}
\end{figure}

\begin{figure}
    \centering
    \includegraphics[width=\linewidth]{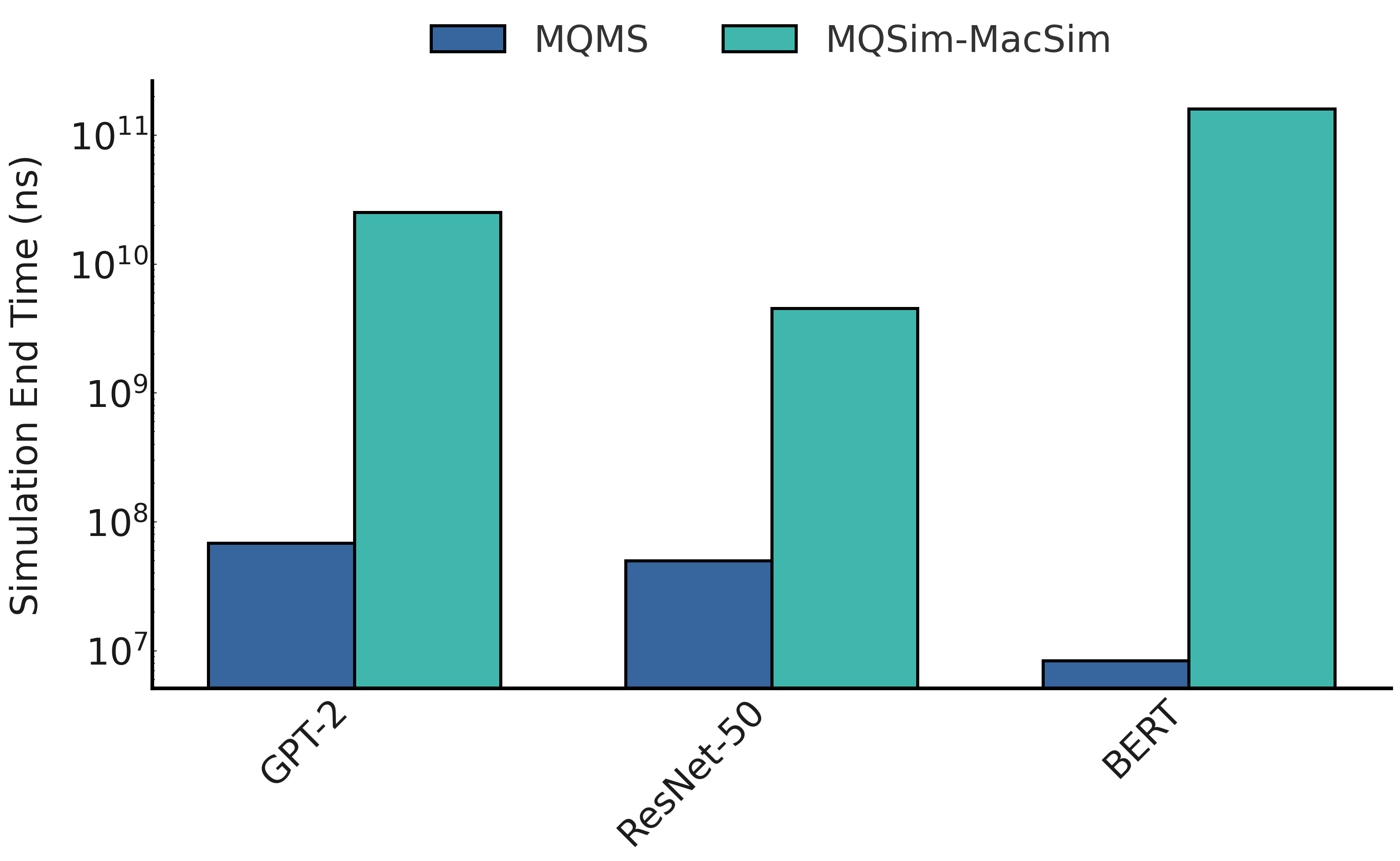}
    \caption{Simulation End Time by Workload}
    \label{fig:llm_sim}
\end{figure}

Figure \ref{fig:llm_iops} compares the IOPS achieved by both simulators for each LLM inference workload. Across all workloads, MQMS exhibits I/O request throughput orders of magnitude beyond that of MQSim-MacSim. This improvement is particularly pronounced for BERT, where the performance gap between MQMS and MQSim-MacSim reaches its maximum, showing an IOPS differential that is an order of magnitude larger than in other workloads. This stark difference stems from BERT's bidirectional architecture, which generates more frequent I/O requests when loading attention weights across multiple layers simultaneously \cite{devlin_2018}. MQMS's ability to exploit plane-level parallelism is particularly effective for such access patterns, as it can distribute these concurrent small requests across multiple planes without the overhead of RMW operations.

Improved IOPS performance catalyzes improvements in both device response time and simulation end time. Device response time, measured as the interval between enqueueing a request in the I/O submission queue and its removal from the I/O completion queue, particularly shows dramatic improvement for MQMS. Figure \ref{fig:llm_resp} shows MQMS maintaining average device response times multiple orders of magnitude lower than MQSim-MacSim across all workloads.

While metrics like cache/memory statistics and page footprint characterize the benchmark behavior, the simulation end time provides a comprehensive view of system performance by capturing the cumulative effect of improved IOPS, reduced latency, and better chip utilization. Figure \ref{fig:llm_sim} indicates MQMS completes simulations up to four orders of magnitude faster than MQSim-MacSim, illustrating how improvements in I/O request servicing latency aggregate to accelerate system-level execution.

\section{Policy Maxima}
The performance characteristics of GPU-SSD systems can vary significantly based on the interplay between scheduling policies and page allocation schemes. Different workloads exhibit unique patterns of I/O requests, memory access, and computational requirements that may be better served by specific policy combinations. By analyzing these policy combinations across various workloads, we identify \textit{policy maxima}---optimal configurations where the combined properties of scheduling and allocation schemes particularly complement specific workload characteristics.

Two primary scheduling policies govern GPU resource allocation in MQMS: round-robin and large chunk. Round-robin scheduling trivially rotates through all active workloads, allocating GPU resources to one kernel from each workload in a circular sequence. Large chunk scheduling emerges as a fallback policy when the number of blocks per kernel falls below the product of block stride and available cores, making fine-grained scheduling inefficient \cite{Kim_2023}. Rather than adhering to strict rotation, this policy processes larger consecutive workload segments before switching. The policy is triggered when $n_{blocks} < s_{block} \times n_{cores}$, where $n_{blocks}$ represents the number of blocks in the current kernel, $s_{block}$ is the configured stride length, and $n_{cores}$ is the number of available GPU cores \cite{Kim_2023}. Large chunk scheduling can also be explicitly selected for workloads where maintaining GPU context is prioritized over fairness, such as in batch processing scenarios with multiple independent tensor operations that share common weights.

Three page allocation schemes are primarily considered in MQMS: CWDP, CDWP (Channel-Die-Way-Plane), and WCDP (Way-Channel-Die-Plane). These schemes determine how logical addresses are mapped to physical flash resources. CWDP prioritizes channel-level parallelism, allocating pages across channels first before considering ways, dies, and planes, which provides lower latencies but potentially underutilizes flash-level resources \cite{Jung}. CDWP modifies this approach by prioritizing die interleaving over way pipelining, offering better flash-level parallelism at the cost of increased system-level resource conflicts \cite{Jung}. WCDP takes a contrasting approach by prioritizing way-level resources first, favoring way pipelining over channel striping, which can achieve improved resource utilization but may incur higher request servicing latencies \cite{Jung}.

\subsection{Evaluation Results}
\begin{figure}[h]
    \centering
    \includegraphics[width=\linewidth]{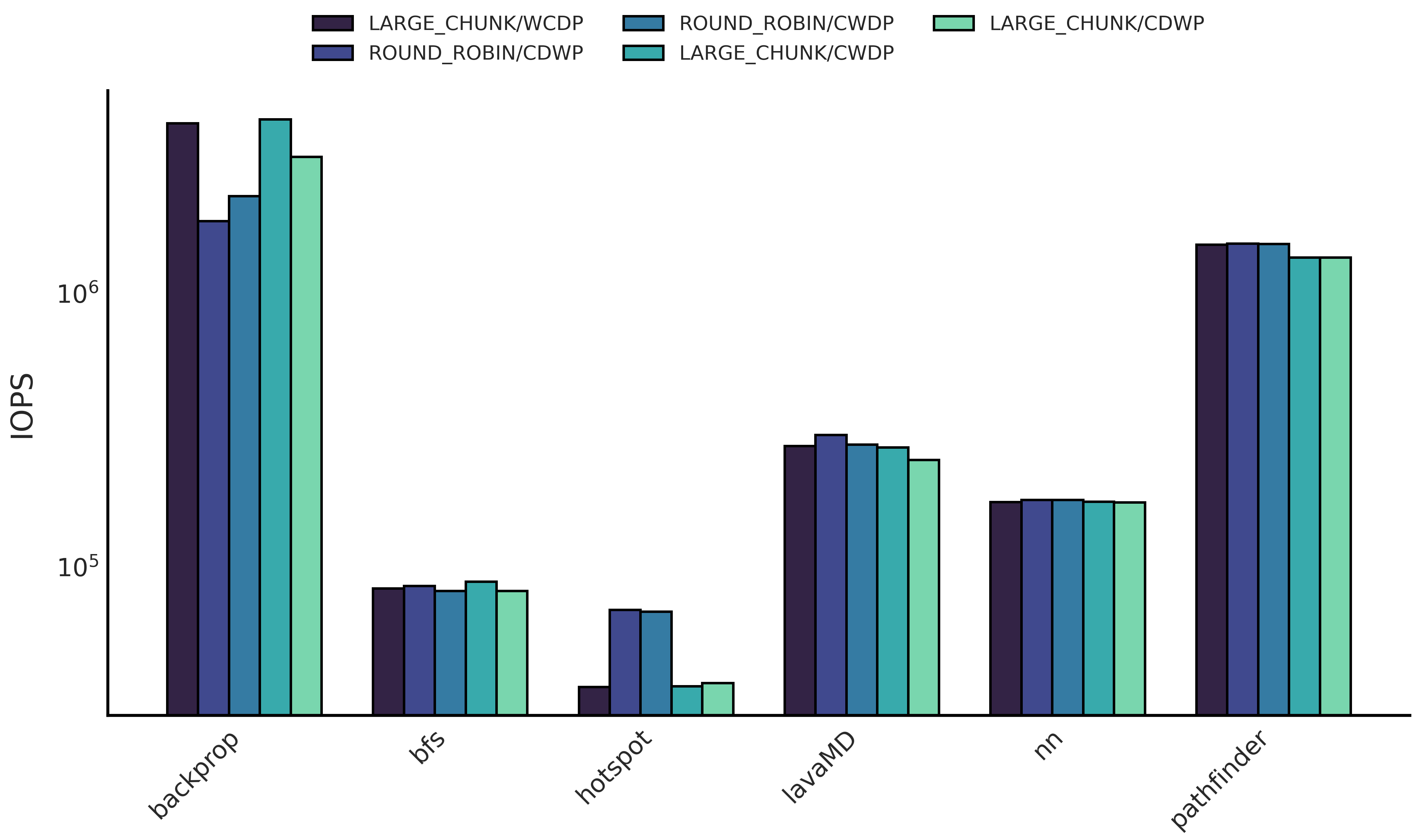}
    \caption{IOPS by Combination}
    \label{fig:maxima_iops}
\end{figure}

\begin{figure}[h]
    \centering
    \includegraphics[width=\linewidth]{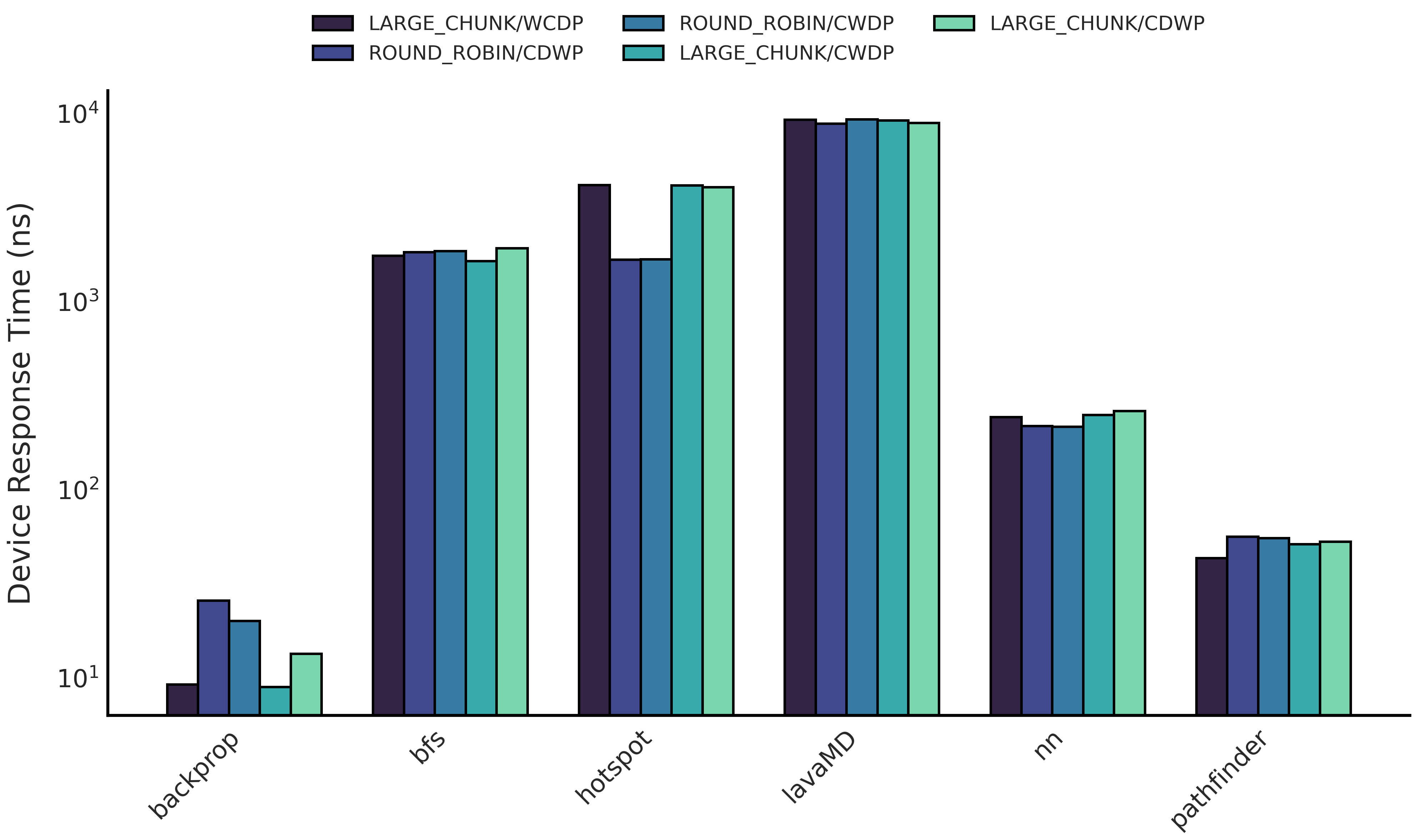}
    \caption{Device Response Time by Combination}
    \label{fig:maxima_resp}
\end{figure}

\begin{figure}[h]
    \centering
    \includegraphics[width=\linewidth]{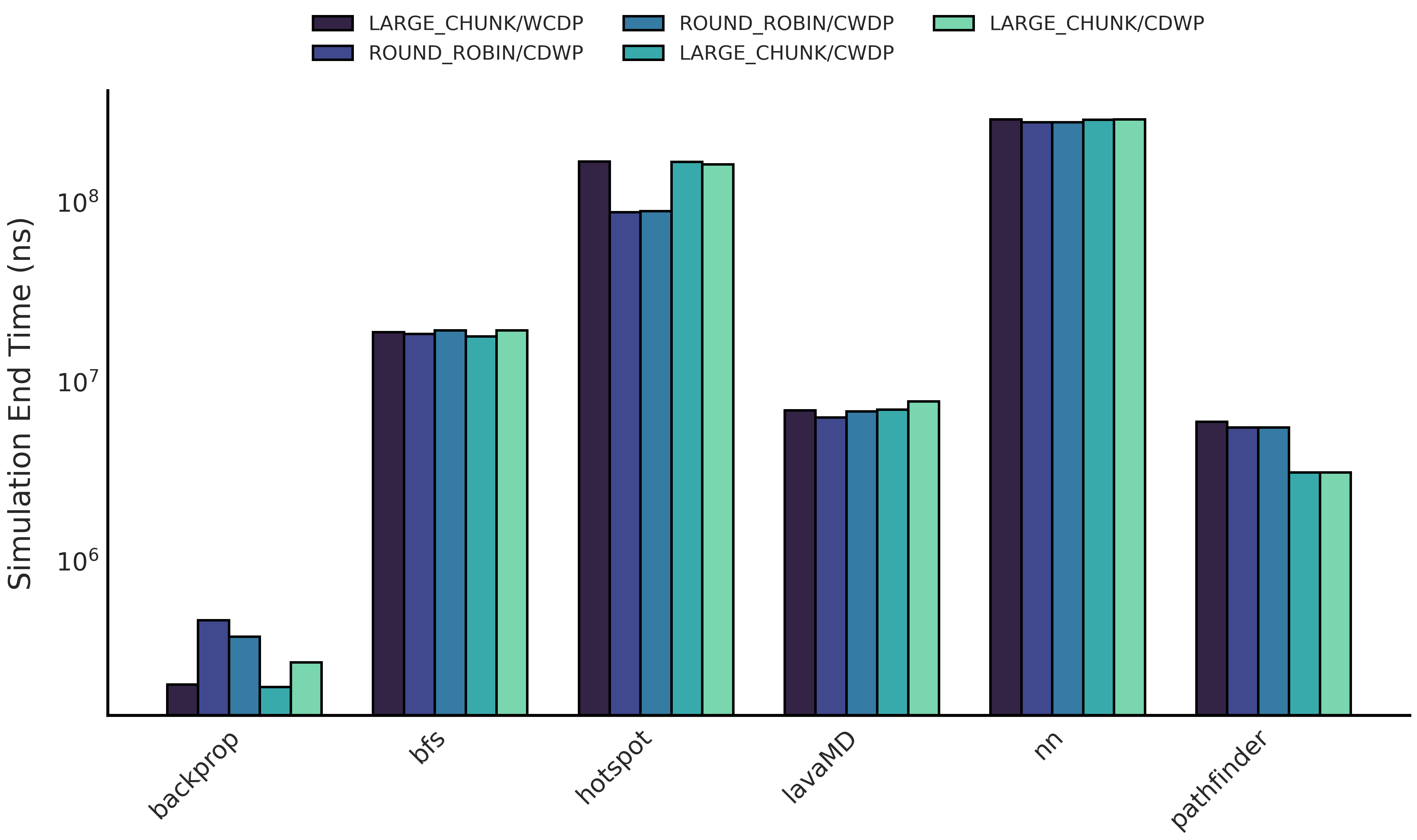}
    \caption{Simulation End Time by Combination}
    \label{fig:maxima_sim}
\end{figure}

Figure \ref{fig:maxima_iops} shows that backprop exhibits the most pronounced performance variation across policy combinations, where large chunk and WCDP delivers a 128\% improvement over round-robin and CDWP due to its regular access patterns and high data locality. In contrast, hotspot shows larger but more erratic variations, with a 92\% performance difference between configurations. Figure \ref{fig:maxima_resp} indicates well-matched policy combinations can dramatically reduce device response times, with backprop showing an ~85\% reduction when using large chunk and CWDP versus round-robin and CDWP. Figure \ref{fig:maxima_sim} reveals similar patterns in simulation end times, with lavaMD achieving a 21\% reduction in execution time under round-robin and CDWP compared to large chunk and WCDP.

\section*{Code Availability}
The source code and additional materials for the MQMS simulator are available on GitHub: \href{https://github.com/ayushgun/mqms}{github.com/ayushgun/mqms}.

\bibliographystyle{abbrv}
\bibliography{refs}

\end{document}